\documentstyle[aps,epsf,tighten,rotate,preprint]{revtex}
\begin{document}

\title{High quality factor measured in fused silica\\
\Large PREPRINT August 11, 2000.}

\author{Steven~D.~Penn, Gregory~M.~Harry, Andri~M.~Gretarsson, Scott~E.~Kittelberger,
Peter~R.~Saulson, John~J.~Schiller, Joshua~R.~Smith, and Sol O. Swords}

\address{Department of Physics, Syracuse University, Syracuse, NY13244-1130}

\maketitle

\begin{abstract}
We have measured the mechanical dissipation in a sample
of fused silica drawn into a rod. The sample was hung from
a multiple-bob suspension, which isolated it from rubbing against
its support, from recoil in the support structure, and
from seismic noise. The quality factor, $Q$,
was measured for several modes with a high value of 57 million
found for mode number 2 at 726 Hz. This result is about a factor 2
higher than previous room temperature measurements. The measured
$Q$ was strongly dependent on handling, with a pristine
flame-polished surface yielding a $Q$ 3--4 times higher than a
surface which had been knocked several times against a copper
tube.
\end{abstract}
\newpage

\section{Introduction}

Several interferometric gravitational wave detectors are currently
being built and commissioned around the
world~\cite{LIGO,VIRGO,GEO,TAMA}, with additional interferometers
planned for the future~\cite{ACIGA,whitepaper}.  The extremely weak
interaction between gravitational waves and laboratory sized masses
requires noise in the interferometer from all sources to be
aggressively minimized.

Thermal excitations represent one of the limiting noise sources
and are expected to be a major constituent of the
LIGO~II~\cite{whitepaper} noise budget in the intermediate
frequency range (roughly 20~Hz to a few hundred Hertz).  Thermal
noise is caused primarily by dissipation in the interferometer
test masses and their suspensions. To minimize thermal noise,
these components must be made from materials with low internal
friction.

Fused silica, an extremely pure form of silicate glass, has been
found to have low internal friction at room
temperature~\cite{Fraser,Bill,Gillespie:Thesis,Lunin,Gretarsson,Litten}.
The current design for LIGO~II specifies that the optic
suspensions be composed entirely of fused silica. Single-crystal
sapphire is being considered for the mirrors themselves, with
fused silica as a fallback material~\cite{whitepaper}.  We have
made dissipation measurements on fused silica samples to help
design the next generation of interferometers.

\section{Experiment}

\begin{figure}
\begin{center}
\epsfxsize=10cm \leavevmode \epsfbox{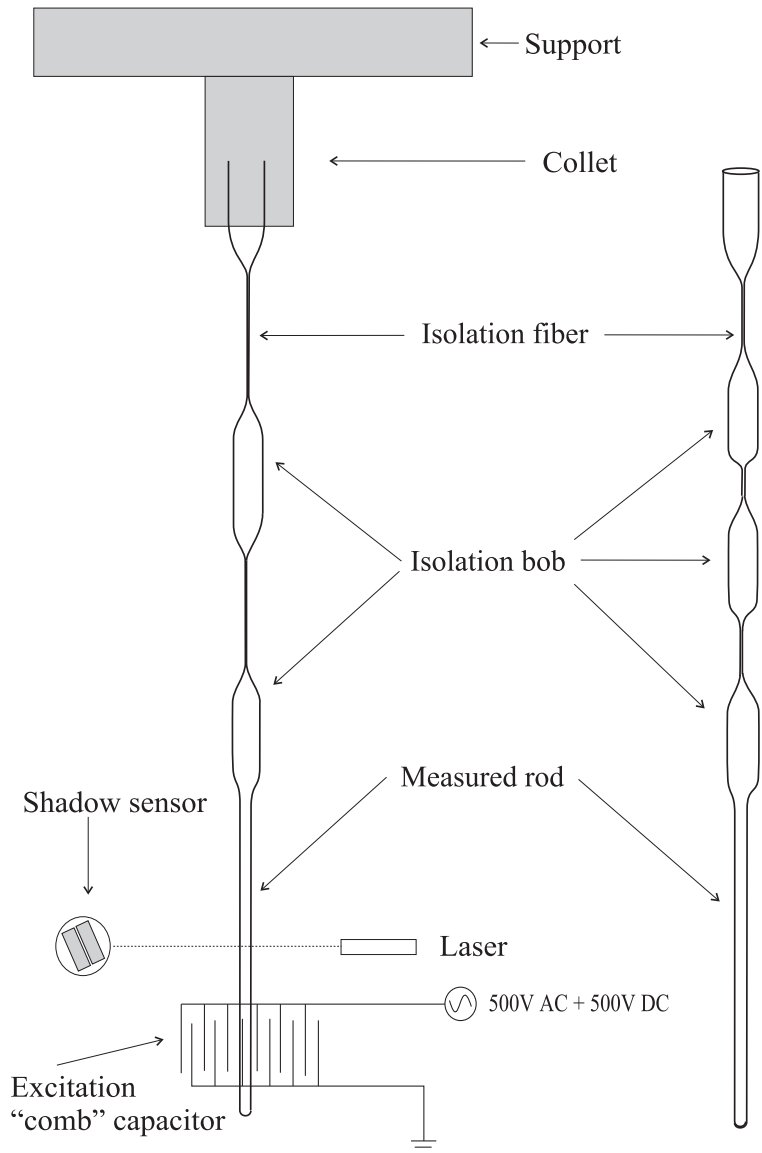}
\end{center}
\caption{Experimental apparatus used to measure $Q$'s.  On the
left, the measured rod is hung beneath the multiple bob suspension.
The top bob is clamped in a collet.  The comb capacitor exciter is
placed near the bottom of the rod, and the beam from the laser
shines higher up on the rod.  The shadow is cast onto a split
photodiode.  On the right, the rod is hung beneath the multiple
bob suspension used in the final round of measurements.}
\label{fig:apparatus}
\end{figure}

Our sample was composed of Suprasil 2 brand fused silica
manufactured by Heraeus Amersil, Inc.  We started with a silica
rod 8~mm in diameter and drew it down to a rod 3.5~mm in diameter and 18.8~cm
in length, using a natural gas flame. A bob of the 8~mm stock was
left attached to one end of the 3~mm sample. A fused silica
suspension was drawn above the sample from the same 8~mm stock.
We made measurements with two different suspensions, both shown
in Figure~\ref{fig:apparatus}.  These suspensions were
distinguished by the number and size of the intermediate
isolation bobs.  The topmost bob was secured in a stainless steel
collet which was bolted to a thick aluminum plate supported by
four aluminum legs.  The whole structure was contained in a
vacuum bell jar which was pumped down below 1~mPa (typically
below 0.1~mPa).

This sample had been previously measured in
1998~\cite{Gretarsson}. The current experimental procedure and
data analysis are nearly identical to those described in that
reference.  The sample was excited at the resonant frequency
using a comb capacitor exciter~\cite{Cadez}. Then, with the
exciter grounded, the sample was allowed to ring down. The
position of the sample was measured using a shadow sensor. The
data was fit to a damped sinusoid from which we extracted the
ringdown time, $\tau$.  Using the mode frequency, $f$, we
calculate $Q = \pi f \tau$.

\begin{figure}
\begin{center}
\epsfxsize=15cm \leavevmode \epsfbox{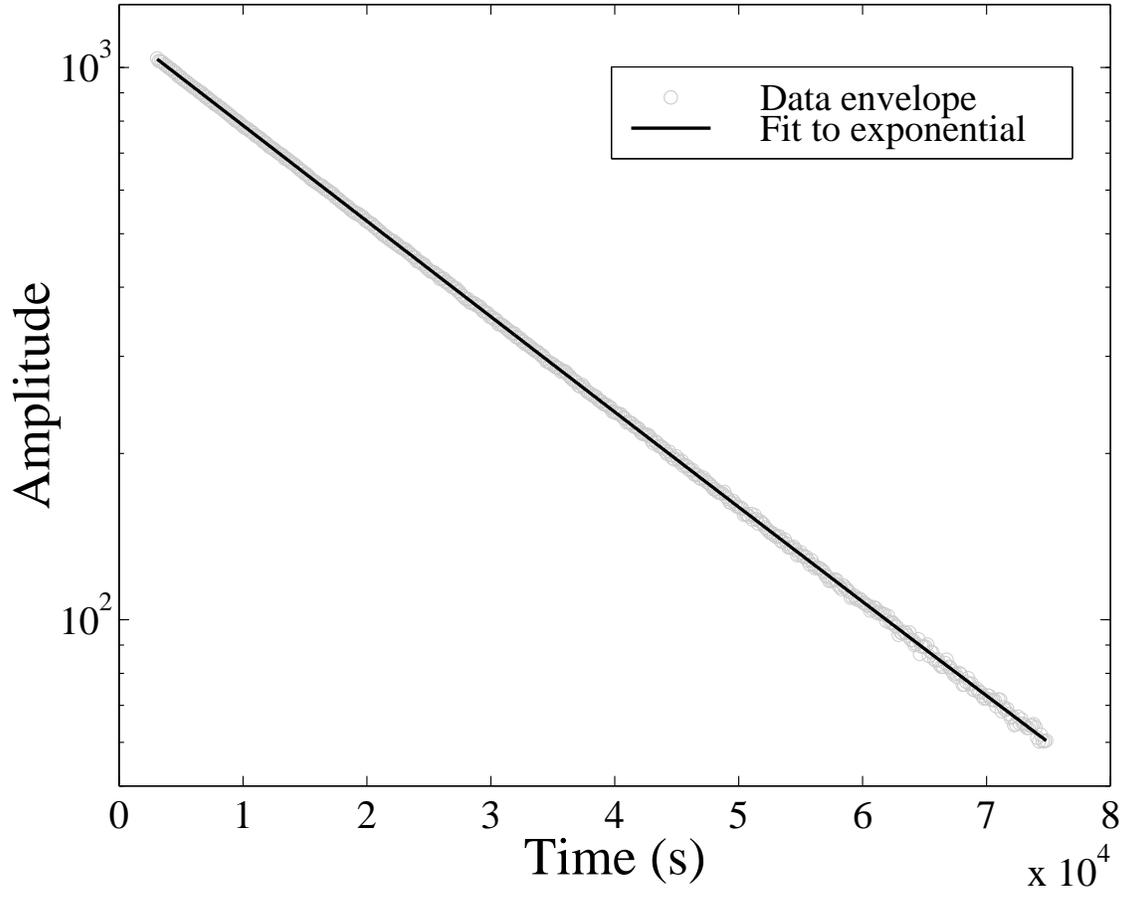}
\end{center}
\caption{Raw data and exponential fit ($Q=5.7\times10^{7}$) for mode 2
immediately after flame polishing the fiber.} \label{fig:typ_data}
\end{figure}

The ring down that yielded our best $Q$ is shown in
Figure~\ref{fig:typ_data}.  The fit to an exponential decay is
excellent.  On the whole, the fit residuals for our data runs were
featureless gaussian noise, with $\sigma_Q/Q \le 1\%$. We observed no evidence
for external excitations (ie.  seismic) in the data, nor did we
account for such excitations in the data analysis.

The scatter in the results of repeated measurements was typically
about 5\%, which is substantially higher than expected given our
gaussian measurement noise (statistical error).  For example, the
fiber mode that yielded our best results had the following scatter:
$5.71\times 10^{7} \pm 9.71\times 10^{4}$, $5.56\times 10^{7} \pm
4.24\times 10^{4}$, $5.50\times 10^{7} \pm 1.45\times 10^{5}$.  We
believe this scatter arises from small excess losses which vary from
run to run.  Possible examples of such losses include rubbing friction
in the clamp, temperature dependence of recoil loss, and eddy current
damping.  These losses would all diminish the measured $Q$ and would
not obey gaussian statistics.  Therefore we quote the highest measured
$Q$ as the best indication of the level of dissipation in a given
sample.


The highest $Q$ found previously for this sample was $2.1 \times
10^{7}$ in both modes 2 and 3~\cite{Gretarsson}. After these
earlier experiments, the sample's diameter was measured with steel
calipers at about 1~cm intervals along its length.  It was then
placed in a borosilicate glass tube and stored at room
temperature on a laboratory shelf for sixteen months.

\begin{figure}
\begin{center}
\epsfxsize=15cm \leavevmode \epsfbox{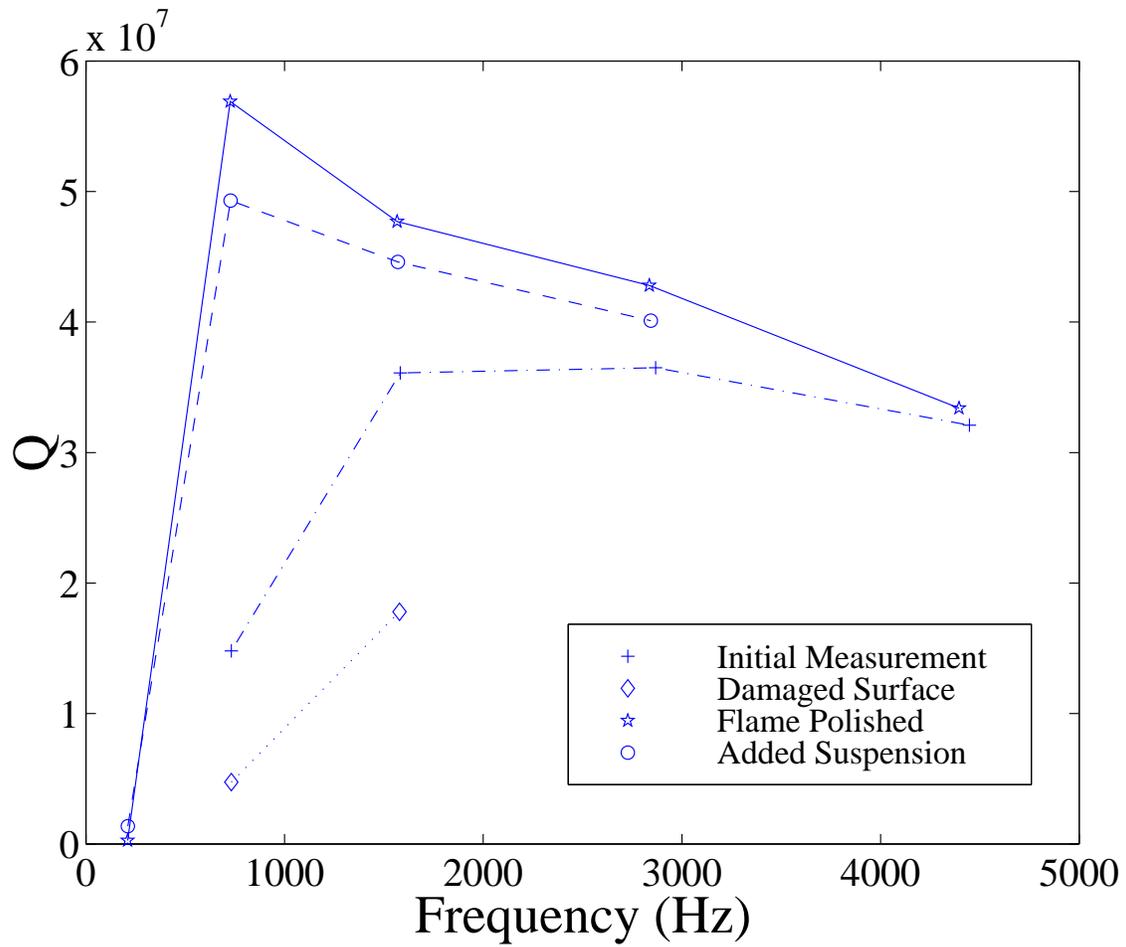}
\end{center}
\caption{Summary of our measured $Q$'s.} \label{fig:sum_data}
\end{figure}

After storage, the sample was removed from the glass tube and hung in
a new vacuum enclosure which is more massive and sits on an optical
table rigidly connected to the floor.  The sample was hung as before
from a collet with the same multiple bob suspension.  Five mechanical
modes of the sample were found using the comb capacitor exciter with a
DC offset of 500~V and an AC signal at the resonant frequency and
amplitude of 500~V. The shadow sensor consisted of a diode laser which
cast the shadow of the silica rod on a split photodiode.  The highest
$Q$'s found for each mode are shown in Table~\ref{table:1} and
Figure~\ref{fig:sum_data}.  For this diameter of fused silica fiber,
thermoelastic damping is negligible at the measured frequencies.

After a first set of measurements, the sample, still hanging from the
multiple bob suspension, was inserted into a 2.8~cm diameter copper tube
in preparation for a separate experiment.  The sample knocked against
the copper tube multiple times during the insertion.  Immediately
following this procedure, the measured $Q$'s were significantly lower
(see Table~\ref{table:1} and Figure~\ref{fig:sum_data}).

We hypothesized that the lower $Q$'s came from an increase in
surface loss~\cite{Gretarsson}, induced by the knocking of the
sample against the copper tube.  To test this hypothesis, the
sample was removed from the upper suspension (all fibers and bobs
above the bob directly connected to the sample) and flame
polished. The polishing was done by fixing the remaining bob in a
clamp and holding a natural gas flame to the sample.  The flame
was held in place long enough so the surface reached the glass
transition temperature.  The flame was then moved back and forth
across the sample so the entire surface reached this point,
although not simultaneously.  The duration of the flame polishing
was 15~minutes.  The sample was then allowed to cool in air until
the bob could be held, another 15~minutes.

The sample was then moved back to the laboratory and immediately
reconnected to the suspension. The original bobs were used, but
new connecting fibers were drawn between all the bobs. These
fibers were drawn to be thinner than on the previous suspension,
with the new fibers being typically $200 \pm 50~\mu$m in
diameter. Extreme care was taken to ensure that the sample
surface was untouched from the time it was flame polished until it
was suspended in the bell jar.

The $Q$'s measured after flame polishing (see Table~\ref{table:1} and
Figure~\ref{fig:sum_data}) showed a marked increase compared to all
previous measurements for this sample.  Indeed these $Q$ values are
nearly twice any previously reported room temperature
measurement~\cite{Luninpersonal} and equal to the highest measurement
of $Q$ in fused silica at any temperature~\cite{Lunintemp}.

The aberrantly low $Q$ value found for mode 1 is believed to be due
to the inability of the suspension to adequately isolate the
sample from modes in the aluminum support structure (recoil
damping) at this low frequency.  The decrease of $Q$ with frequency
for modes 2--5 is not understood. It may indicate the frequency
dependence of an internal loss mechanism such as an Arrhenius
loss peak~\cite{Zener}, or it may be due to acoustic radiation traveling up
the suspension.

To test the effect of the suspension on the sample $Q$'s, we
removed the sample from the suspension and redrew the suspension
with one additional bob.  The fibers were again drawn to be $200
\pm 50~\mu$m thick.  In order to keep the entire suspension plus
sample roughly the same height, two smaller bobs had to be
substituted for the larger central bob.  These new bobs were also
of Suprasil 2 fused silica.  The redrawing of the suspension and
repositioning of the sample in the vacuum chamber was once again
done with great care so the sample's surface was not touched. The
measured $Q$'s using this new suspension are shown in
Table~\ref{table:1} and Figure~\ref{fig:sum_data}.

\section{Discussion}

The $Q$'s we measured of the flame-polished fused silica rod
are the highest ever observed for fused silica at room
temperature.  The previous highest measurement at room
temperature, $3 \times 10^{7}$, was obtained by
Lunin~\cite{Lunin,Luninpersonal} using a ``wine glass''
resonator. The damaged surface layer of this resonator was
removed using an unpublished chemical treatment.


The observed dependence of $Q$ on surface condition is generally
consistent with prior work~\cite{Lunin,Gretarsson}. However, the
results of this experiment do supersede the quantitative
conclusion of Ref.~\cite{Gretarsson}, as the $\phi (=1/Q)$ value
lies significantly below the curve shown in Figure~6 of that
reference.  Thus, that paper's predicted value for the bulk
dissipation of fused silica, $\phi_{\mathrm{bulk}} = 3 \times
10^{-8}$, is too high and instead should be considered indicative
of the level of excess loss in these prior measurements.

\begin{figure}
\begin{center}
\epsfxsize=10cm \leavevmode \epsfbox{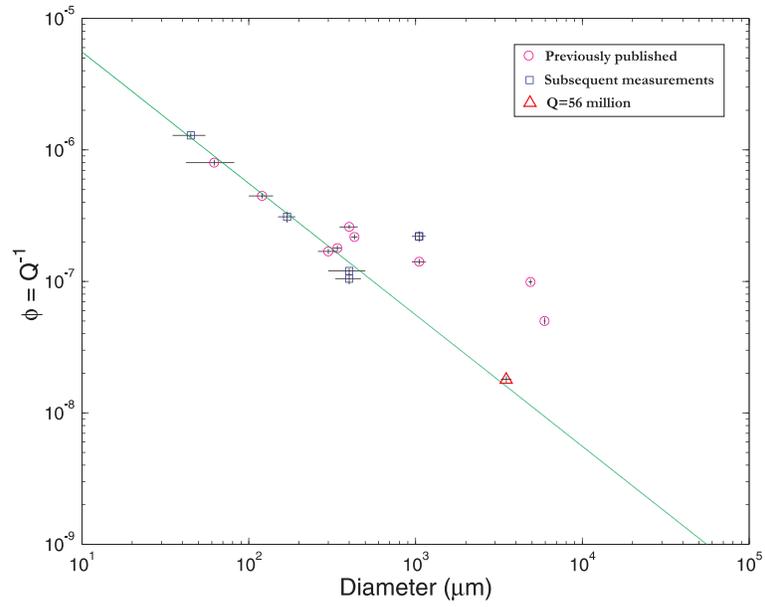}
\end{center}
\caption{Data from Figure~6 of \protect\cite{Gretarsson} and subsequent measurements. Current
result is marked by a triangle. The line has slope -1, which is consistent with
loss proportional to the surface-to-volume ratio.} \label{fig:stov}
\end{figure}

In Figure~\ref{fig:stov} we have plotted our current data along with
the data from Ref.~\cite{Gretarsson}.  We have also included some more
recent data taken in the same way as in that paper.  Using the
theoretical model presented in Ref.~\cite{Gretarsson}, we have plotted
the line indicating the dependence of $\phi$ in the surface limited
regime.  All the data, including our recent measurements, are limited
by this line, suggesting that even the highest measured $Q$'s are
limited by surface loss.  If that were the case, then the bulk
$Q$ for fused silica would be greater than $5.7 \times 10^{7}$. The present
data are not sufficient to determine $\phi_{\mathrm{bulk}}$
using our theoretical model.  Further research in this area is clearly
warranted.


It is interesting to compare our results here with the results of
Lunin~\cite{Lunintemp}, who, using a fused silica hemispherical
resonator at 8400~Hz, found the $Q$ to peak at 105~C at a value of
55~million.  If the loss mechanisms that dominates Lunin's results are
Arrhenius processes with similar activation energies, then at a lower
resonant frequency, the $Q$ would peak at a lower temperature.  It is
possible that at 726~Hz, the peak is near room temperature.  We are
currently conducting an experiment to measure the temperature
dependence of $Q$ at these lower resonant frequencies.


To test whether our results were limited by suspension losses, we
added an additional isolation bob in our fiber suspension. Except
for mode 1, the $Q$'s did not change much, which indicates that
modes 2--5 are not suspension limited and that their losses arise
from internal mechanisms.  With the additional isolation bob, the
$Q$ of mode 1 increased from $3 \times 10^{5}$ to $4 \times
10^{6}$. This increase indicates that mode 1 was suspension
limited by the initial multiple bob suspension. The fact that the $Q$
for mode 1 is still well below that of the higher modes, suggests
that even with the new suspension mode 1 is still suspension
limited.

If these $Q$ values could be obtained in a test mass of a
gravitational wave interferometer, the thermal noise would be
significantly lower than currently anticipated for silica
optics~\cite{whitepaper}. The thermal noise from the mirrors'
internal modes expressed as a spectral density of gravitational
wave strain is given by~\cite{Liu}
\begin{displaymath}
S_{h}\left(\omega\right) = \frac{4 k_B T}{\omega L^2}
\frac{1-\sigma^2}{\sqrt{2 \pi} E r_0} \phi\left(\omega\right),
\nonumber
\end{displaymath}
where $k_B$ is Boltzmann's constant, $T$ is the temperature, $L$ is
the arm length of the interferometer, $\sigma$ is Poisson's ratio, $E$
is Young's modulus, $r_0$ is the width of the laser beam, and $\phi$
is the loss angle.  In this frequency band, we have assumed that $\phi
= 1/Q$~\cite{Gretarsson}.  If fused silica optics with $Q = 3.0\times
10^{7}$ were used in LIGO II, their thermal noise would dominate the
noise budget over a couple of octaves centered on $f = 100$~Hz.  The
thermal noise associated with $Q = 5.7\times 10^{7}$ is almost
$\sqrt{2}$ lower, a substantial improvement.


Actually realizing this lower thermal noise in the mirrors' internal modes
could prove quite challenging. The dependence of $Q$ on surface
condition could rule out high $Q$'s in a polished mirror.  Even
obtaining the $3 \times 10^7$ estimated in the LIGO~II
design~\cite{whitepaper} may prove difficult because the mirror
surfaces of the test masses and beam splitters must be polished
and coated.  There is evidence that these coatings may
limit achievable thermal noise~\cite{coating}. We are currently
conducting experiments to explore this issue.

It is unlikely that our current results could help reduce the low
frequency noise in the LIGO II design.  While fused silica ribbons are
planned for the final stage of the test mass suspension, their thermal
noise, which should be dominated by surface loss, is already expected to
be so low~\cite{Glasgow_ribbon,ribbon} that it is comparable to the
radiation pressure noise.

\section*{Acknowledgments}

We would like to thank John Chabot for his talent and skill in drawing
and flame polishing the fused silica rod; Lou Buda, Charlie Brown, and
Lester Schmutzler for their machining efforts; and Adam Schuman and
Brian Gantz for their help.  We also thank William Startin for
originally suggesting an experiment that led to these results.  This
work was supported by Syracuse University and National Science
Foundation Grant No.  PHY-9900775.

\begin{table}
\begin{tabular}{rrrrr}

Mode Number & Q (i) & Q (ii) & Q (iii) & Q (iv) \\
\hline
1 & & & $0.264 \times 10^6 $ & $1.36 \times 10^6 $ \\
2 & $14.8 \times 10^6$ & $4.74 \times 10^6$ & $57.1 \times 10^6$ & $49.3 \times 10^6 $  \\
3 & $36.1 \times 10^6$ & $17.8 \times 10^6$ & $47.7 \times 10^6$ & $44.6 \times 10^6 $  \\
4 & $36.5 \times 10^6$& & $42.8 \times 10^6$ & $40.1 \times 10^6 $  \\
5 & $32.1 \times 10^6$& & $33.4 \times 10^6$ &
\end{tabular}
\caption{Quality factors of fused silica sample: (i) after storage,
(ii) after knocking against a copper tube, (iii) after flame
polishing, and (iv) with an additional bob in the suspension.}
\label{table:1}
\end{table}

\pagebreak[4]

\pagebreak[4]

\pagebreak[4]

\pagebreak[4]


\begin{thebibliography}{00}

\bibitem{LIGO}A.~Abramovici {\it et al}, Science, {\bf 256} 325
(1992)

\bibitem{VIRGO}A.~Giazotto {\it et al.}, Nucl. Instrum. Methods A
{\bf 289},  518 (1988).

\bibitem{GEO}K.~Danzmann {\em et al.}, GEO 600: Proposal for a 600~m
laser interferometric gravitational wave antenna,
Max-Planck-Institut f\"ur Quantenoptik Report 190, Garching
Germany (1994).

\bibitem{TAMA}K.~Tsubono, in {\em Gravitational Wave Experiments},
proceedings of the First Edoardo Amaldi Conference, World
Scientific, 112, (1995).

\bibitem{ACIGA}D.~E.~McClelland, M.~B.~Gray, D.~A.~Shaddock, B.~J.~Slagmolen,
S.~M.~Scott, P.~Charlton, B.~J.~Whiting, R.~J.~Sandeman,
D.~G.~Blair, L.~Ju, J.~Winterflood, D.~Greenwood, F.~Benabid,
M.~Baker, Z.~Zhou, D.~Mudge, D.~Ottaway, M.~Ostermeyer,
P.~J.~Veitch, J.~Munch, M.~W.~Hamilton, C.~Hollitt, in
{\em Gravitational Waves}, proceedings of the
Third Edoardo Amaldi Conference, ed. S. Meshkov,
Melville NY: American Institute of Physics, 140-149 (2000),
(http://131.215.125.172/info/paperindex/pdf/McClelland.pdf).

\bibitem{whitepaper} {\em LSC White Paper on Detector Research and
Development}, (http://www.ligo.caltech.edu/docs/T/T990080-00.pdf).

\bibitem{Fraser}D.~B.~Fraser, J. Appl. Phys. {\bf 39}, 5868 (1968); {\bf 41}, 6 (1970).

\bibitem{Bill}W.~J.~Startin, M.~A.~Beilby, and P.~R.~Saulson, Rev. Sci. Instr.
{\bf 69} 3681 (1998).

\bibitem{Gillespie:Thesis}A.~D.~Gillespie, PhD. thesis
(California Institute of Technology, 1995) Unpublished.

\bibitem{Lunin}B.~S.~Lunin, S.~N.~Torbin, M.~N.~Danachevskaya, and
I.~V.~Batov, Moscow University Chemistry Bulletin {\bf 35}, 24
(1994), Allerton Press, Inc.

\bibitem{Gretarsson}A.~M.~Gretarsson and G.~M.~Harry, Rev.~Sci.~Instr., {\bf
70}, 4081 (1999), (http://xxx.lanl.gov/abs/physics/9904015).

\bibitem{Litten}E.~J.~Loper, D.~D.~Lynch, and K.~M.~Stevenson, IEEE PLANS
(Position Location and Navigation Symposium) Record, Caesar's
Palace, Las Vegas, Nevada, November 4-7, 1986, pp. 61-64, Table 2.

\bibitem{Cadez}A.~Cadez and A.~Abramovici, J.~Phys E: Sci. Instr. {\bf 21} 453 (1988).

\bibitem{Luninpersonal}B.~S.~Lunin, Chemistry Department, Moscow
State University (private communication, May 1997).

\bibitem{Lunintemp}B.~S.~Lunin and S.~N.~Torbin, {\em Vestnik
Moskovskogo Universiteta}, ser. 2, {\em Khimia}, {\bf 41} 93
(2000).  English version in press.

\bibitem{Zener} C.~M.~Zener, {\em Elasticity and Anelasticity in Metals},
Univ.~of~Chicago Press, Chicago, 1948.

\bibitem{Liu}Y.~T.~Liu and K.~S.~Thorne, submitted to
Phys.~Rev.~D, (http://xxx.lanl.gov/abs/gr-qc/0002055).

\bibitem{coating}A.~M.~Gretarsson, G.~M.~Harry, S.~D.~Penn,
P.~R.~Saulson, J.~J.~Schiller, W.~J.~Startin, in {\em Gravitational
Waves}, proceedings of the Third Edoardo Amaldi Conference, ed.  S.
Meshkov, Melville NY: American Institute of Physics, 306-312 (2000),
(http://131.215.125.172/info/paperindex/pdf/Gretarsson.pdf).

\bibitem{Glasgow_ribbon}S.~Rowan, R.~Hutchins, A.~McLaren, N.~A.~Robertson,
S.~M.~Twyford, and J.~Hough, Phys.~Lett.~A, {\bf 227} 153 (1997).

\bibitem{ribbon}A.~M.~Gretarsson, G.~M.~Harry, P.~R.~Saulson,
S.~D.~Penn, W.~J.~Startin, J.~Hough, S.~Rowan, G.~Cagnoli, Phys.
Lett.~A, {\bf 270}, 108 (2000), (http://xxx.lanl.gov/abs/gr-qc/9912057).

\end{thebibliography}
\end{document}